# Degeneration of kernel regression with Matern kernels into low-order polynomial regression in high dimension

Sergei Manzhos[1], Manabu Ihara

School of Materials and Chemical Technology, Tokyo Institute of Technology, Ookayama 2-12-1, Meguro-ku, Tokyo 152-8552 Japan

## Abstract

Kernel methods such as kernel ridge regression and Gaussian process regressions with Matern type kernels have been increasingly used, in particular, to fit potential energy surfaces (PES) and density functionals, and for materials informatics. When the dimensionality of the feature space is high, these methods are used with necessarily sparse data. In this regime, the optimal length parameter of a Matern-type kernel tends to become so large that the method effectively degenerates into a low-order polynomial regression and therefore loses any advantage over such regression. This is demonstrated theoretically as well as numerically on the examples of six- and fifteen-dimensional molecular PES using squared exponential and simple exponential kernels. The results shed additional light on the success of polynomial approximations such as PIP for medium size molecules and on the importance of orders-of-coupling based models for preserving the advantages of kernel methods with Matern type kernels or on the use of physically-motivated (reproducing) kernels.

## 1 Introduction

Machine learning is increasingly used in physical chemistry, materials science, and computational chemistry with applications ranging from fitting potential energy surfaces (PES) to construction of exchange-correlation and kinetic energy functionals to prediction of materials properties from descriptors of chemical composition and structure.[1–11] Kernel methods such as kernel ridge regression (KRR) and Gaussian process regressions (GPR) are widely used for these purposes.[12–14] Here, we consider only numeric features (inputs) $\boldsymbol{x}$. One then fits a target function $f(\boldsymbol{x}), \boldsymbol{x} \in R^D$ to a set of $M$ samples $f_n = f(\boldsymbol{x}^{(n)}), n = 1, \ldots, M$. GPR and KRR represent the target function in the same form

$$f(\boldsymbol{x}) = \boldsymbol{f}^* \boldsymbol{K}^{-1} \boldsymbol{f} \equiv \sum_{n=1}^{M} k(\boldsymbol{x}, x^{(n)}) c_n$$

(1.1)

where $k(\boldsymbol{x}, \boldsymbol{x}')$ is the kernel, $\boldsymbol{f}^*$ is a row vector with elements $f_n^* = k(\boldsymbol{x}, \boldsymbol{x}^{(n)})$, and $\boldsymbol{f}$ is the vector of all known values $f_n$. $\boldsymbol{K}$ is the variance-covariance matrix,

$$\boldsymbol{K} = \begin{pmatrix} k(\boldsymbol{x}^{(1)}, \boldsymbol{x}^{(1)}) + \delta & \cdots & k(\boldsymbol{x}^{(1)}, \boldsymbol{x}^{(M)}) \\ \vdots & \ddots & \vdots \\ k(\boldsymbol{x}^{(M)}, \boldsymbol{x}^{(1)}) & \cdots & k(\boldsymbol{x}^{(M)}, \boldsymbol{x}^{(M)}) + \delta \end{pmatrix}$$

(1.2)

---

[1] E-mail: manzhos.s.aa@m.titech.ac.jp

The (hyper)parameter $\delta$ is often added on the diagonal of $\boldsymbol{K}$ to improve stability (for example, when $\boldsymbol{K}$ is ill-conditioned) and generalization property of the approximation; it is a regularization parameter.[12] Eq. (1) is a linear regression over basis functions $k_n(\boldsymbol{x}) = k(\boldsymbol{x}, \boldsymbol{x}^{(n)})$ built from the kernel functions, with coefficients $\boldsymbol{c} = \boldsymbol{K}^{-1} \boldsymbol{f}$. Rather than introducing $\delta$ into a square, $M{\times}M$, matrix $\boldsymbol{K}$ (which is the typical view of KRR and GPR), one can stabilize the solution by using fewer basis functions $N$ than sample points $M$. The matrix $\boldsymbol{K}$ is then of size $M{\times}N$ and the coefficient $\boldsymbol{c}$ are determined in the least squares sense from its pseudoinverse.[15]

Matern type kernels are most widely used. They have the form[16]

$$k(\boldsymbol{x}, \boldsymbol{x}') = \sigma^2 \frac{2^{1-\nu}}{\Gamma(\nu)} \left(\sqrt{2\nu} \frac{L}{l}\right)^{\nu} K_{\nu}\left(\sqrt{2\nu} \frac{L}{l}\right)$$

(1.3)

where $\Gamma$ is the gamma function, and $K_\nu$ is the modified Bessel function of the second kind. $L$ is the distance between points; often the Euclidean distance is used $L = |\boldsymbol{x} - \boldsymbol{x}'|$. The prefactor $\sigma^2$ is usually taken as the variance of $\{f_n\}$ and is omitted in the following as it is fully correlated with $\delta$ and otherwise can be accounted for by $\boldsymbol{c}$. In most applications, $\nu$ is fixed at a value resulting is one of simple, well understood functions: as $\nu \to \infty$, one obtains the widely used square exponential kernel also called the RBF (radial basis function) kernel:

$$k(\boldsymbol{x}, \boldsymbol{x}') = \exp\left(-\frac{L^2}{2l^2}\right)$$

(1.4)

at $\nu = 1/2$, one obtains a simple exponential kernel

$$k(\boldsymbol{x}, \boldsymbol{x}') = \exp\left(-\frac{L}{l}\right)$$

(1.5)

and for $\nu = 3/2$ and $5/2$, one obtains Matern3/2 and Matern5/2 kernels, respectively, that are products of an exponential and a polynomial term in $L/l$. In this work, we will therefore consider the RBF and the simple exponential kernel. The key hyperparameter of these kernels is the length parameter $l$ (Eqs. 4-5 are written for an isotropic kernel which is reasonable when the data are normalized or scaled in all dimensions; anisotropic versions with different $l_i, i = 1, \ldots, D$ can also be used).

The GPR view of kernel regression relies on a view of Eq. 1.1 as a mean of a Gaussian process and on the use of kernel functions of the Matern type that can be viewed as covariance functions between different datapoints. These kernels are traditionally considered as local as they monotonically decay with the distance between datapoints $L$. When the dimension $D$ of the feature space becomes high, the property of locality may be lost, as was recently demonstrated in Ref. [17]. This has to do with two factors: (i) the relevant measure of locality of a basis function (which is the role played by the kernel in Eq. 1.1) in many applications (notably in computational chemistry) is not the amplitude but quadrature;

in high $D$, functions that are local in low $D$ are no longer local by this measure (the standard example is the Gaussian function, about 68% of whose quadrature is concentrated within one standard deviation from the mean when $D = 1$ but only about 2% when $D = 10$). (ii) in high $D$, data density is necessarily sparse and cannot be significantly increased by adding more data in any practical sense due to the so-called curse of dimensionality.[18] This causes the optimal hyperparameter $l$ to become large and much larger than then characteristic length scale of the data in any dimension. For example, when fitting with GPR the six-dimensional PES of $H_2CO$ in the spectroscopically relevant region to several thousand normalized data, the optimal (minimizing the error on a large test set) $l$ was about 5; when fitting the fifteen-dimensional PES of $UF_6$, the optimal $l$ was about 30.[19] As $D$ increases, the values of the matrix $\boldsymbol{K}$ are distributed closer and closer to 1. For example, in the case of the PES of $UF_6$, with the optimal $l$, all values of elements of $\boldsymbol{K}$ were larger than about 0.95 – the locality of the kernel was lost.[17] This has incidence elsewhere, e.g. the loss of advantage of a multi-$\zeta$ basis which is intrinsically tied to locality. In contrast, in the case of PESs of $H_2O$ (3D) and $H_2CO$ (6D), values of elements of $\boldsymbol{K}$ noticeably smaller than 1 were abundant.[17] When fitting the Nikkei 225 stock market index in a $D = 2346$ dimensional feature space, to data scaled on unit cube, the optimal $l$ was more than 500.[20] It was demonstrated that when $D \rightarrow \infty$, GPR approximation necessarily collapses when the distribution of the data on which the model is used is any different from the distribution of the training data, which it practically always is, unless also $l \rightarrow \infty$.[20,21] In the case of the 2346-dimensional fit, any advantage over simple ($f(\boldsymbol{x}) = \sum_{d=0}^{D} x_d$) linear regression was lost. In the case of the 15-dimensional $UF_6$ potential, the advantage over simple linear regression is maintained (simple linear regression cannot form a good baseline model of a PES as it behaves parabolically around the equilibrium geometry).

Then $l$ is very large (compared to the characteristic length scale of the data in any dimension), the kernel loses resolution in any one dimension. Here, we show that is effectively works in an asymptotic region with the approximation of Eq. 1.1 degenerating into a simple low-order polynomial regression. The implication of this is that in high $D$ and with sparse data, there is no advantage of GPR or KRR with *full-dimensional* Matern type kernels over simple polynomial regression e.g. of the PIP (permutationally invariant polynomial) kind, which has proven to be quite successful for medium size molecules.[22–26] To recover the superior expressive power of kernel regression with this types of kernel, one would need to lower the dimensionality of the kernel, for example by using orders-of-coupling expansions.[19,27–30] Alternatively, one could resort to specially designed non-local kernels, as in the RKHS (reproducing kernel Hilbert space) method.[31–34]

## 2   Methods

The calculations are performed in Matlab with a home-made code using the rectangularized version of GPR.[15] We use rectangular formulation to avoid dealing with the selection of the $\delta$ parameter on the diagonal of $\boldsymbol{K}$ and focus on the kernel behavior with respect to the length parameter. The matrix $\boldsymbol{K}$ is then of the form

$$\boldsymbol{K} = \begin{pmatrix} k(\boldsymbol{x}^{(1)}, \boldsymbol{x}^{(1)}) & \cdots & k(\boldsymbol{x}^{(1)}, \boldsymbol{x}^{(N)}) \\ \vdots & \ddots & \vdots \\ k(\boldsymbol{x}^{(M)}, \boldsymbol{x}^{(1)}) & \cdots & k(\boldsymbol{x}^{(M)}, \boldsymbol{x}^{(N)}) \end{pmatrix}$$

(2.1)

where $N$ is the number of basis functions and $M$ the number of training datapoints. A random subset of $N$ points out of $M$ is used, as the ratio $M/N$ is kept at 1.4 based on the results of Ref. [15]. The function approximation is then $f(\boldsymbol{x}) = \boldsymbol{f}^* \boldsymbol{K} \backslash \boldsymbol{f}$ using the Matlab's "\" operator (which uses the smallest rank solution), which we find to be more accurate in Matlab for large $l$ (when near-linear dependency of the basis arises) than the *pinv* function (which uses the lowest norm solution) in $f(\boldsymbol{x}) = \boldsymbol{f}^* pinv(\boldsymbol{K}) \boldsymbol{f}$ , that was used in Ref. [15]. The data are normalized and therefore isotropic RBF and simple exponential kernels (Eqs. 1.4-5) are used.

When $l \to \infty$, the RBF kernel behaves as

$$k(\boldsymbol{x}, \boldsymbol{x}') = \exp\left(-\frac{L^2}{2l^2}\right) = \exp\left(-\frac{|\boldsymbol{x} - \boldsymbol{x}'|^2}{2l^2}\right) \approx \prod_{d=1}^{D}\left(1 - \frac{(x_d - x_d')^2}{2l^2}\right)$$

$$(2.2)$$

and

$$k(\boldsymbol{x}, \boldsymbol{x}') = \exp\left(-\frac{d}{l}\right) \approx 1 - \frac{d}{l}$$

$$(2.3)$$

This results in low-order polynomial approximations: with Eq. 2.2,

$$f(\boldsymbol{x}) = (2l^2)^{-D} \sum_{n=1}^{N} c_n \prod_{d=1}^{D}\left(\left[2l^2 - \left(x_d^{(n)}\right)^2\right] - x_d^2 + 2x_d x_d^{(n)}\right)$$

$$(2.4)$$

with the highest order terms of the form $\prod_{d=1}^{D} x_d^{p_d}$ where $p_d \leq 2, \forall d$ and $\sum_{d=1}^{D} p_d = 2D$. This form is expected to work well for near-parabolic functions. The simple exponential kernel with the Euclidean distance results in a less tractable asymptotics math, so to illustrate the point of devolution to a low-order polynomial approximation, we will use the version with the city block distance (in which case $k(\boldsymbol{x}, \boldsymbol{x}') = \prod_{d=1}^{D} exp\left(-\frac{|x_d - x_d'|}{l}\right) \approx \prod_{d=1}^{D}\left(1 - \frac{|x_d - x_d'|}{l}\right)$). We have confirmed in numeric tests that both the Euclidean and city block distance fitting results are similar. With the city block distance, when $l \to \infty$,

$$f(\boldsymbol{x}) = l^{-D} \sum_{d=1}^{D} c_n \prod_{d=1}^{D}\left(l - \left|x_d - x_d^{(n)}\right|\right)$$

$$(2.5)$$

which is a piecewise low-order polynomial approximation with bilinear highest order terms of the form $\prod_{d=1}^{D} x_d^{p_d}$ where $p_d \leq 1, \forall d$ and $\sum_{d=1}^{D} p_d = D$.

Polynomial approximations can also be obtained with dot product based kernels. With the standard (first order) dot product kernel $k(\boldsymbol{x}, \boldsymbol{x}') = \chi + \boldsymbol{x}\boldsymbol{x}'$, one obtains for Eq. 1.1

$$f(\boldsymbol{x}) = \left( \chi \sum_{n=1}^{N} c_n \right) + \sum_{d=1}^{D} \left( \sum_{n=1}^{N} c_n x_d^{(n)} \right) x_d = b_0 + \sum_{d=1}^{D} b_d x_d$$

(2.4)

i.e. a simple linear regression. With $k(\boldsymbol{x}, \boldsymbol{x}') = \chi + (\boldsymbol{x}\boldsymbol{x}')^2$, one obtains

$$f(\boldsymbol{x}) = \left( \chi \sum_{n=1}^{N} c_n \right) + \sum_{d=1}^{D} \left( \sum_{n=1}^{N} c_n \left( x_d^{(n)} \right)^2 \right) x_d^2 + \sum_{d \neq d'}^{D} \left( \sum_{n=1}^{N} c_n x_d^{(n)} x_{d'}^{(n)} \right) x_d x_{d'}$$

$$= b_0 + \sum_{d=1}^{D} b_d^{(2)} x_d^2 + \sum_{d \neq d'}^{D} b_{dd'} x_d x_{d'}$$

(2.5)

i.e. a parabolic approximation. A combination of these, i.e. $k(\boldsymbol{x}, \boldsymbol{x}') = \chi + \boldsymbol{x}\boldsymbol{x}' + (\boldsymbol{x}\boldsymbol{x}')^2$, obtains a second order polynomial approximation with both linear and bilinear terms. Note that the choice of $\chi$ and of coefficients at $\boldsymbol{x}\boldsymbol{x}'$ and $(\boldsymbol{x}\boldsymbol{x}')^2$ terms in the kernel is immaterial as long as $N$ is large enough, as they result in the same $b_0, b_d$ etc. by the adjustment of $c_n$. Similarly, by using higher order powers of $\boldsymbol{x}\boldsymbol{x}'$ in the kernel, one obtains higher order polynomial approximations of $f(\boldsymbol{x})$.

When kernel regression devolves into a low-order polynomial approximation at high $l$ in particular applications, we should then observe that the fit quality becomes similar to that with the above dot product based kernels when $l$ is optimal (best for the test set error). This is what we demonstrate below. While, when fitting the six-dimensional PES of $H_2CO$, much better results are obtained with the RBF kernel than with the dot product based kernels, indicating the ability of kernel regression to account for anharmonicity and coupling terms of complex form, when fitting the 15-dimensional PES of $UF_6$, there is no advantage of the RBF (and even less so of the exponential) kernel vs a low-order polynomial regression.

We use the same datasets sampling the PESs of $H_2CO$ and $UF_6$ in the spectroscopically relevant region as were used in Refs. [19,27] (and are available from those references). The $H_2CO$ dataset contains 120,000 points sampling an analytic PES in bond coordinates with energies ranging from 0 (equilibrium geometry) to 17,500 cm$^{-1}$. The $UF_6$ dataset contains 50,000 points computed with DFT with energies ranging from 0 (equilibrium geometry) to 5,000 cm$^{-1}$ and with data given in normal mode coordinates. The reader is referred to Refs. [19,27] for more details about the data including PES scans along different variables.

For both datasets, we show results with 5,600 training and 20,000 test points. We use a large test set to reliably gauge the global quality of the fit. We confirmed that the conclusions do not change when using other numbers of points. This is because the advantages of the Matern kernel are well observed for $H_2CO$ with this number of points, and while theoretically increasing the number of training points in $UF_6$ fits would eventually allow profiting from the higher expressive power with Matern kernels (vs simple polynomial regression), it is not practically feasible. Indeed, the entire dataset of 50,000 available DFT points corresponds to an effective density of sampling of only 2.06 data per degree of freedom (DOF), not much higher than the 1.78 data/DOF with the 5600 training points. It would take 2.3 billion datapoints to achieve a density of sampling equivalent to that of the $H_2CO$ PES (about 4.2 data/DOF) in 15D.

## 3    Results

### 3.1    6D: sufficient density of sampling allows profiting from the superior expressive power of Matern type kernels

In Figure 1, top row, we show the results of the fits of the $H_2CO$ PES with RBF and exponential (city block distance) kernels with optimal length parameters, which are about 5 for RBF and 15 for the exponential kernels. These values of $l$ are on the order of the range of the (normalized) features, which is about [-2.3…-2.6, 2.6…3.8] depending on the feature, preserving the resolution of the kernel in any dimension. To this corresponds the distribution of the values of the elements of $\boldsymbol{K}$ which includes many values significantly smaller than 1 (see Ref. [17]). The rmse (root mean square error) and correlation coefficient values for training and test sets are indicated on the plots. There is minor variability of the numbers due to particular random selection of train and test points, but it is insignificant on the scale of the numbers. The RBF kernel clearly outperforms.

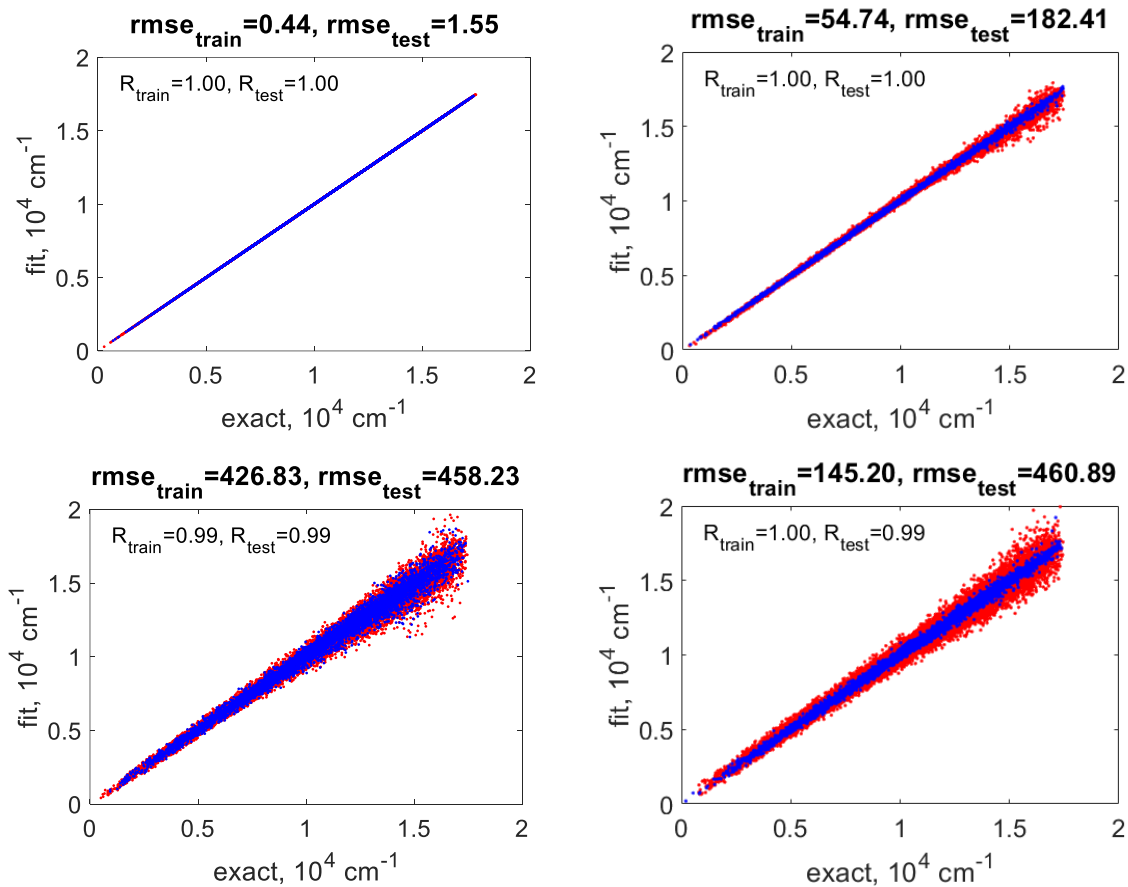

Figure 1. Top row: correlation plot between the exact (reference) and fitted $H_2CO$ potential energy values when fitting with RBF (left) and exponential (right) kernels with optimal length parameters (5 and 15, respectively). Bottom row: same with artificially large length parameters (50 and 2000, respectively). Blue: training points, red: test points. Where test points are not visible, they visually overlap with the training points. The rmse and correlation coefficient values for training and test sets are indicated on the plots.

Figure 1 also shows, in the bottom row, the results of the fits when the length parameter is set to large values, 50 for RBF and 2000 for the exponential kernel. At these $l$ values, the values of all elements of $\boldsymbol{K}$ are higher than 0.99. Not surprisingly, a significant deterioration of the fit quality is observed. The fit remains reasonable even at very large values of $l$ with correlation coefficient of about 0.99 (the visible spread of the data in the figure somewhat overstates the error as many points concentrated along the diagonal visually overlap). The calculation can become numerically unstable when $l$ reaches values much higher than these.

In Figure 2, we show the results of the fit using first and second order dot product based kernels, $k(\boldsymbol{x}, \boldsymbol{x}') = \sum_{p=0}^{P}(\boldsymbol{x}\boldsymbol{x}')^p$ for different $P$. One must go much beyond the quadratic approximation to approach, albeit not quite, the accuracy achieved with an RBF kernel (at $P$ = 6). This test also demonstrates the significant effect of anharmonic contributions: the second order polynomial approximation ($P = 2$) achieves a mediocre rmse on the order of 400 cm$^{-1}$, which is much larger than the test error obtainable with the optimal-length RBF kernel and *which is similar to that achieved with the RBF kernel when l is made to be very high*, as suggested by Eq. 2.4. Overall, in this case, a sufficiently high density of sampling allows profiting from the superior expressive power of kernel regression with Matern type kernels compared to simple polynomial models. This is indeed a sampling density issue: for example, when using only 140 training points, the optimal $l$ of the RBF kernel becomes about 15 (with all elements of $\boldsymbol{K}$ becoming larger than 0.9) and the test set rmse is on the order of 700-800 cm$^{-1}$ (depending on a particular draw of random training points). The same accuracy is then obtained with $k(\boldsymbol{x}, \boldsymbol{x}') = \sum_{p=0}^{3}(\boldsymbol{x}\boldsymbol{x}')^p$ i.e. with a 3$^{rd}$ order polynomial model, as can be appreciated from Figure 3. This sampling density (corresponding to about 2.3 data/DOF) is much lower than is typically used (and easily obtained) in relatively low-dimensional problems like this one. This contrasts with the 15-dimensional case of UF$_6$ considered next, where even with many thousands of training points the equivalent sampling density is barely above 2 data/DOF.

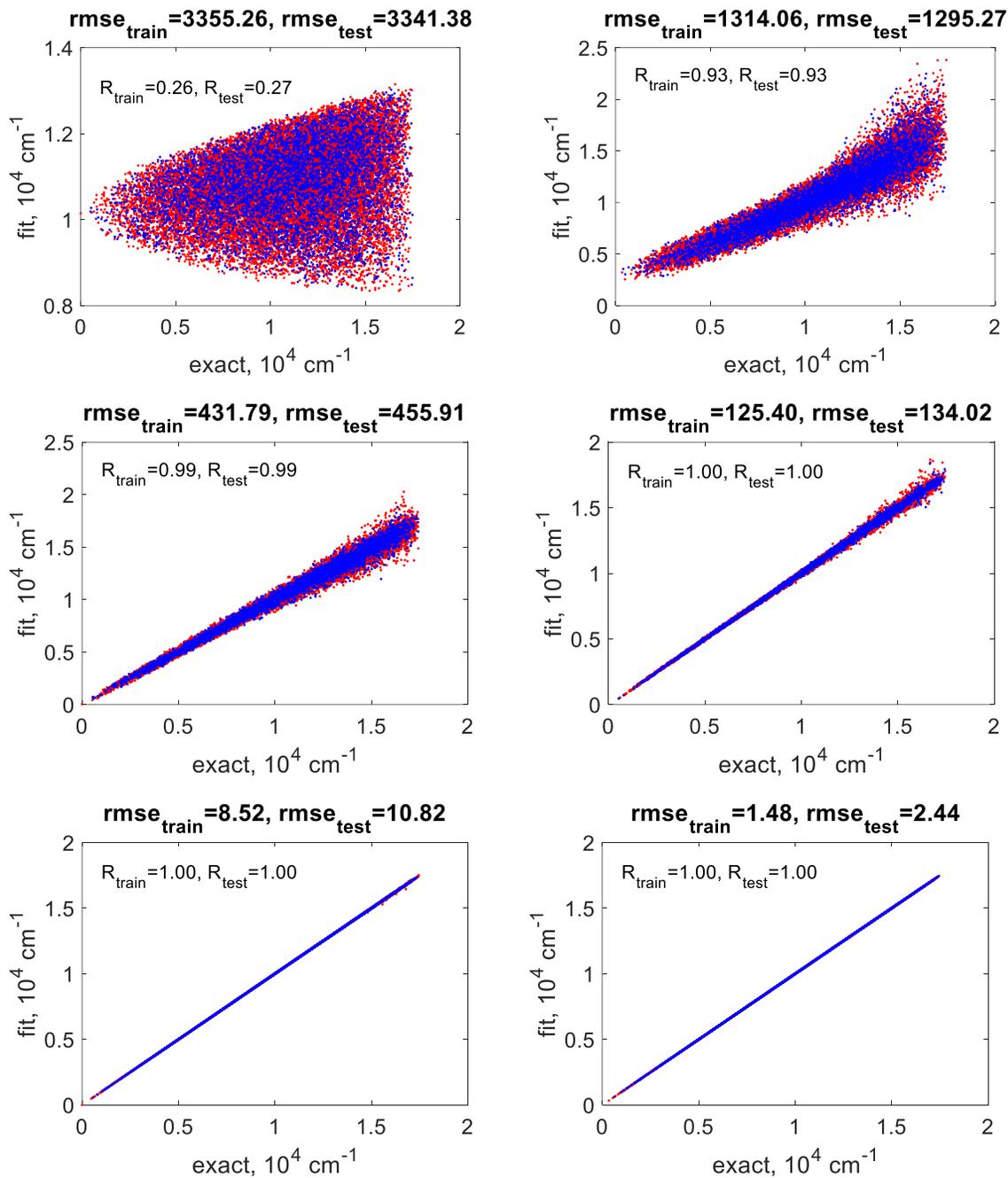

Figure 2. Correlation plot between the exact (reference) and fitted $H_2CO$ potential energy values when fitting using kernels $k(\boldsymbol{x}, \boldsymbol{x}') = \sum_{p=0}^{P}(\boldsymbol{x}\boldsymbol{x}')^p$ with different $P$ from 1 to 6 (left to right and top to bottom). Blue: training points, red: test points. Where test points are not visible, they visually overlap with the training points. The rmse and correlation coefficient values for training and test sets are indicated on the plots.

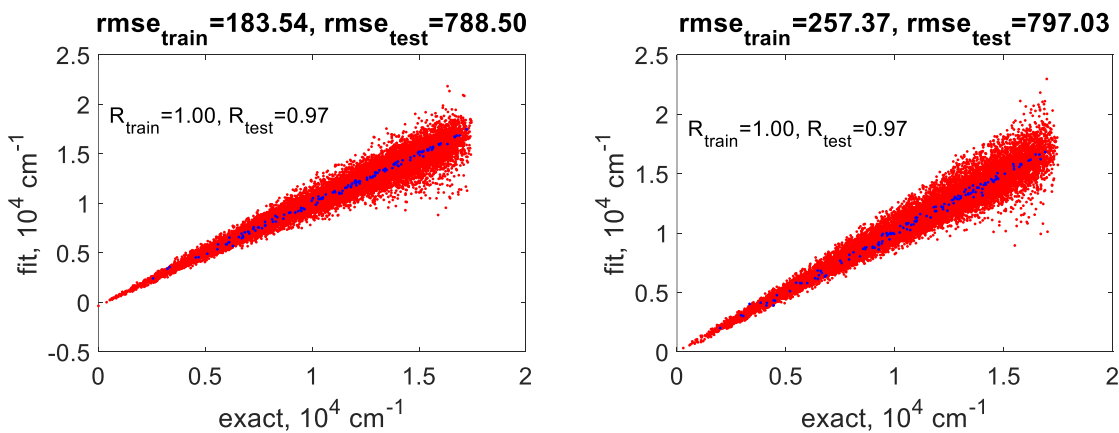

Figure 3. Left: correlation plot between the exact (reference) and fitted $H_2CO$ potential energy values when fitting (left) with RBF kernels with an optimized length parameter (15) and (right) with $k(\boldsymbol{x}, \boldsymbol{x}') = \sum_{p=0}^{3} (\boldsymbol{x}\boldsymbol{x}')^p$ kernel. Blue: training points, red: test points. The rmse and correlation coefficient values for training and test sets are indicated on the plots.

### 3.2 15D: degeneration of kernel regression with Matern kernels to a low-order polynomial approximation

In Figure 4, we show the results of the fits of the $UF_6$ PES with RBF and exponential (city block distance) kernels with optimal length parameters, which are about 30 for RBF and also 30 for the exponential kernel, i.e. much large than the range of the (normalized) features, which is about [-2, 2] for all features. Lower values of $l$ result in a worse test set rmse. To this corresponds the distribution of the values of the elements of $\boldsymbol{K}$ with the RBF kernel which are all > 0.95 (see also Ref. [17]). Due to the use of the city block distance with the exponential kernel, the optimal $l$ with it is not larger than for the RBF kernel in this higher-dimensional case. The relatively smaller $l$ with the exponential kernel using city block distance results in many values of the elements of $\boldsymbol{K}$ being on the order of 0.5. Using Euclidean distance results in an optimal $l$ of about 200 (there is rather little change in fit quality within the range of $l$ of 100-300 with this kernel) and best test set rmse of about 160 cm$^{-1}$ i.e. similar to the city block distance version. In this case all elements of $\boldsymbol{K}$ are larger than 0.9.

The rmse and correlation coefficient values for training and test sets are indicated on the plots. The RBF kernel clearly outperforms in this case too when the length parameter is optimal. The bottom panels of the figure also show the results the length parameter is set to even larger values, 300 (RBF) and 2000 (exponential kernel). At these $l$ values, the values of all elements of $K$ are higher than 0.999 with the RBF kernel and higher than 0.98 with the exponential kernel. Not surprisingly, a significant deterioration of the fit quality is observed, which nevertheless remains reasonable.

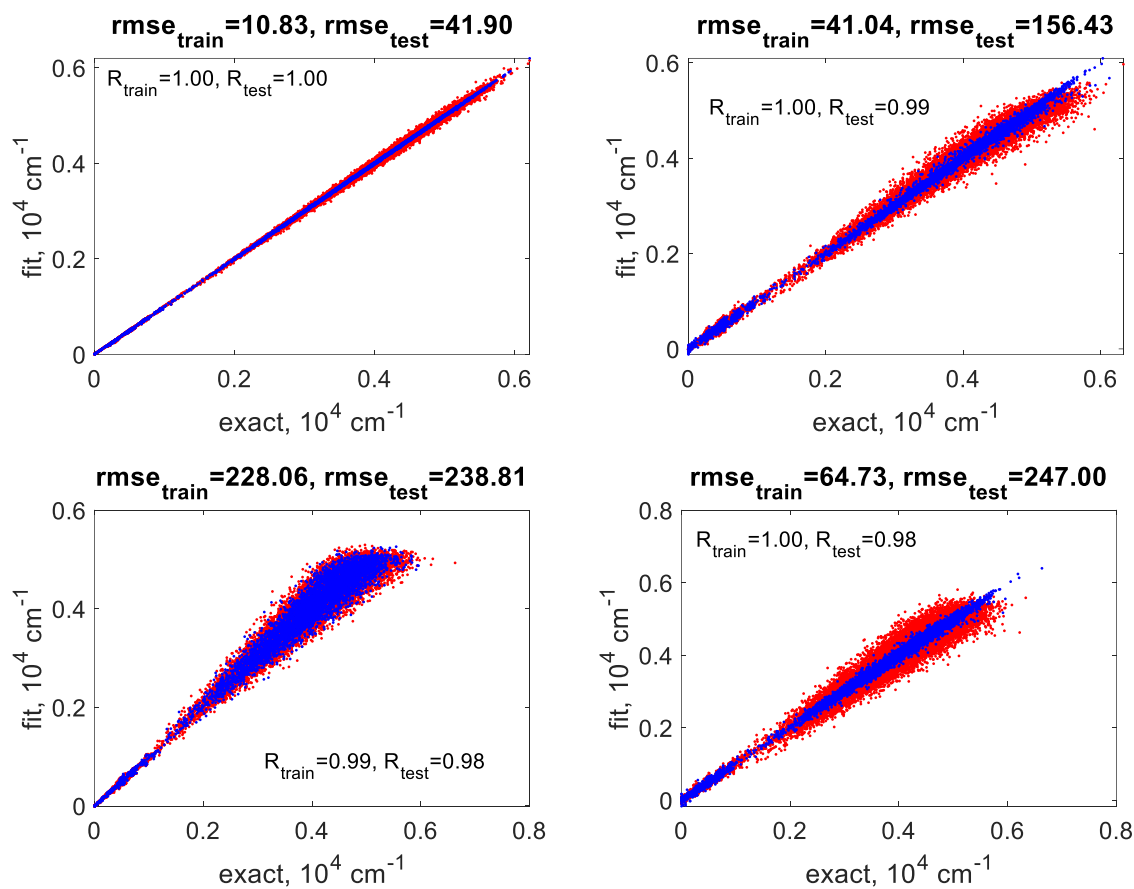

Figure 4. Top row: correlation plot between the exact (reference) and fitted UF6 potential energy values when fitting with RBF (left) and exponential (right) kernels with optimal length parameters (30). Bottom row: same with artificially large length parameters (300). Blue: training points, red: test points. Where test points are not visible, they visually overlap with the training points. The rmse and correlation coefficient values for training and test sets are indicated on the plots.

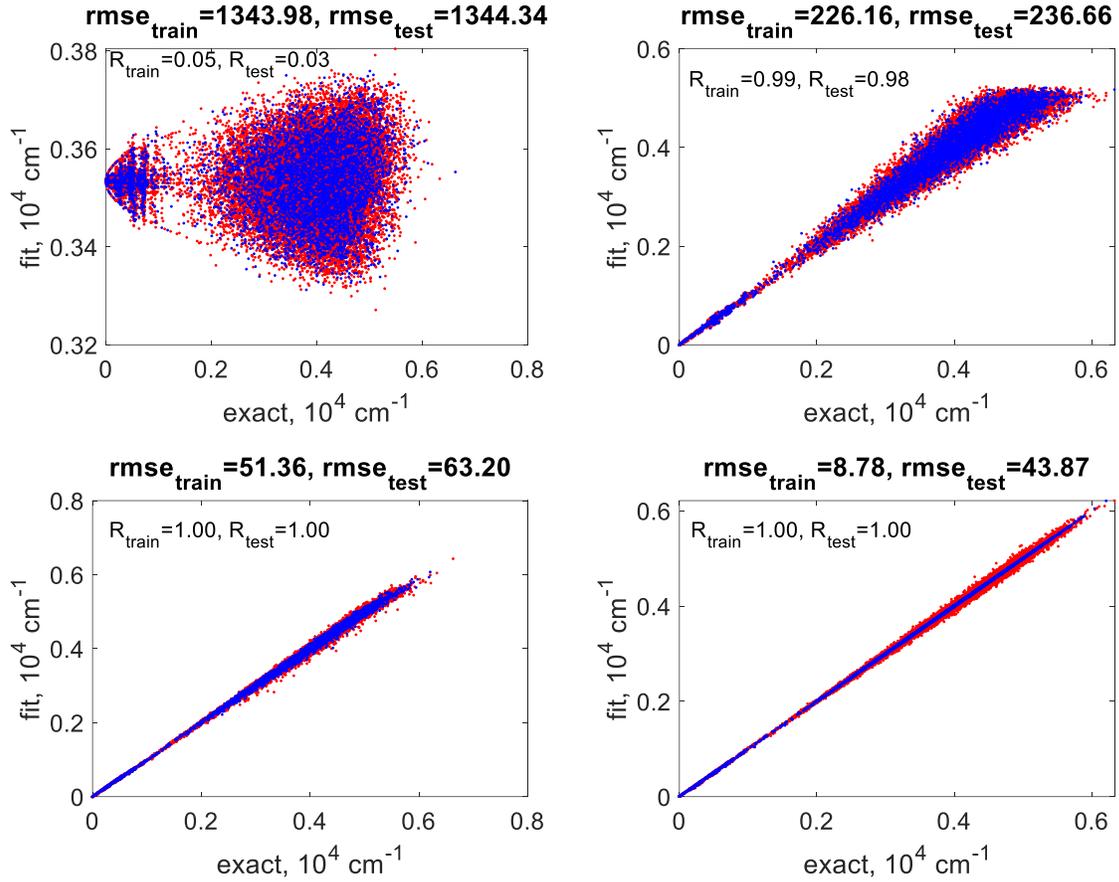

Figure 5. Correlation plot between the exact (reference) and fitted $UF_6$ potential energy values when fitting using kernels $k(\boldsymbol{x}, \boldsymbol{x}') = \sum_{p=0}^{P}(\boldsymbol{x}\boldsymbol{x}')^p$ with different $P$ from 1 to 4 (left to right and top to bottom). Blue: training points, red: test points. Where test points are not visible, they visually overlap with the training points. The rmse and correlation coefficient values for training and test sets are indicated on the plots.

In Figure 5, we show the results of the fit using first and second order dot product based kernels, $k(\boldsymbol{x}, \boldsymbol{x}') = \sum_{p=0}^{P}(\boldsymbol{x}\boldsymbol{x}')^p$ for different $P$. With a 4th order polynomial model ($P = 4$) one obtains the same quality of the model as the best fit with a Matern kernel (test errors on the order of 40 cm⁻¹). There is no improvement possible for $P > 4$ because the number of polynomial terms exceeds the number of training points (as the number of monomials $\prod_{d=1}^{D} x_d^{p_d}$ with $\sum_{d=1}^{D} p_d = \wp$ scales as $\binom{D + \wp}{D}$). The quadratic model ($P = 2$) achieves a similar fit quality to that obtained with Matern kernels with large (much larger than optimal) length parameters (test errors on the order of 240 cm⁻¹). A better fit quality than that of a quartic model is not obtainable from the data and therefore, contrary to the six-dimensional $H_2CO$ case, the superior expressive power of GPR or KRR with Matern kernels cannot be utilized.

## 4 Discussion and conclusions

When machine learning in high-dimensional feature spaces, the sampling density with training data is bound to be sparse due to the curse of dimensionality. In this case, the optimal length parameter of a Matern type kernel, such as the most widely used square exponential (RBF) kernel, will tend to be larger than the characteristic length scale of input features. This leads to a degeneration of the kernel regression to a low-order polynomial model and loss of the superior expressive power of kernel regression with Matern kernels. Here, we demonstrated this effect theoretically as well as computationally by comparing kernel regressions of a six-dimensional PES of formaldehyde and 15-dimensional PES of $UF_6$. In theory, this issue can be resolved by increasing data density. In practice, however, this is not feasible in sufficiently high dimensionality due to the exponential growth in the required data, the so-called curse of dimensionality.

In high-dimensional applications, including construction of high dimensional potential energy surfaces, when using kernel regression with full-dimensional Matern kernels, one thus may not be able to get an advantage over low-order polynomial models. This situation might explain while the state of the art in PES construction today prominently features PIP (permutationally invariant polynomials), which have proven to be quite successful in producing PESs of medium size molecules almost on an industrial scale.[22–26]

Is degeneration of GPR or KRR with a Matern kernel into a low-order polynomial approximation in high $D$ a bad thing? – Not necessarily. The result of such a regression is informative in the sense that it is an indication that with a given dataset the sampling density is insufficient to fully make use of the superior expressive power or kernel regressions with Matern type kernel. If the fit quality is sufficiently good regardless, then this is also an indication that one switch to a low-order polynomial approximation for that particular application, which has its benefits including simplicity, robustness and ease of integration or differentiation.

The degeneration of the kernel regression with Matern kernels can be avoided if the dimensionality of the kernel is lowered. The same property of the sparsity of data that leads to this degeneration can be used to palliate it: when the data density is low, only low-order coupling terms may be recoverable,[35] and then order-of-coupling representations[29,36,37] are useful. The terms of those representations are low-dimensional and when they are represented with KRR or GPR,[19,27,28,38] the superior expressive power of Matern kernels can be preserved. The use of additive kernels,[39] which results in order-of-coupling representations[19,29] is also then recommended. Alternatively, one could use non-Matern types of kernels, such as non-local, physically motivated kernels often employed with the RKHS method.[31–34] Neural networks as representations that use in general non-local, non-direct product basis functions[40,41] are also attractive in this sense. Combinations of NN with one-dimensional Matern type kernels that serve to express the neuron activation functions are also possible.[30,38]

# 5 Acknowledgements

This work was supported by JST-Mirai Program Grant Number JPMJMI22H1, Japan. S. M. thanks Tucker Carrington for discussions.

# 6 Data availability statement

The datasets are available in Refs. [19,27]. The Matlab code can be obtained from the authors upon reasonable request.